\documentclass{aa} 
\usepackage{graphicx}
\usepackage{natbib}
\usepackage[varg]{txfonts}
\newcommand{\ltsima}{$\; \buildrel < \over \sim \;$}
\newcommand{\simlt}{\lower.5ex\hbox{\ltsima}}
\newcommand{\gtsima}{$\; \buildrel > \over \sim \;$}
\newcommand{\simgt}{\lower.5ex\hbox{\gtsima}}

\newcommand{\cgs}{${\rm erg~cm}^{-2}~{\rm s}^{-1}$} 
\newcommand{\lum}{\rm erg~s$^{-1}$}
\newcommand{\lx}{\rm $L_{2-10~keV}$}

\def\lesssim{\mathrel{\hbox{\rlap{\hbox{\lower4pt\hbox{$\sim$}}}\hbox{$<$}}}}
\def\gtrsim{\mathrel{\hbox{\rlap{\hbox{\lower4pt\hbox{$\sim$}}}\hbox{$>$}}}}

\def\arcmin{\hbox{$^\prime$}}
\def\arcsec{\hbox{$^{\prime\prime}$}}
\def\micron{\hbox{$\mu$m}}
\def\mjy{\hbox{$\mu$Jy}}

\def\kms{\rm km~s$^{-1}$}
\def\ab1450{$AB_{1450(1+z)}$}

\def\xray{\hbox{X-ray}}

\def\oiii{\hbox{[O\ {\sc iii}}]}

\def\09104{IRAS~09104$+$4109}
\def\I09104{I09104}

\def\msun{M$_{\odot}$}
\def\edd_ratio{$\log\ L_{\rm bol}/L_{\rm Edd}$}

\def\Nh{{N$_{\rm H}$}}
\def\l58{{$(\lambda L_{\lambda})_{\mbox{{\rm \scriptsize 5.8\micron}}}$}}
\def\lmir2{{$(\lambda L_{\lambda})_{\mbox{{\rm \scriptsize 12.3\micron}}}$}}
\def\s1{{S$_{\mbox{{\rm \scriptsize 3.6\micron}}}$}}
\def\irac2{{S$_{\mbox{{\rm \scriptsize 4.5\micron}}}$}}
\def\f3{{S$_{\mbox{{\rm \scriptsize 5.8\micron}}}$}}
\def\mic8{{S$_{\mbox{{\rm \scriptsize 8\micron}}}$}}
\def\f24{{F$_{\mbox{{\rm \scriptsize 24\micron}}}$}}

\def\chandra{{\it Chandra\/}}

\def\heao1{{\it HEAO-1\/}}

\def\spitzer{{\it Spitzer\/}}
\def\herschel{{\it Herschel\/}}

\def\scuba{{\it SCUBA\/}}
\def\laboca{{\it LABOCA\/}}
\def\aztec{{\it AzTEC\/}}
\def\alma{{\it ALMA\/}}
\def\vla{{\it VLA\/}}
\def\vlba{{\it VLBA\/}}
\def\atca{{\it ATCA\/}}

\def\xmm{{XMM-{\it Newton\/}}}

\def\nature{{Nature}}
\def\science{{Science}}
\usepackage{color}

\begin{document} 

\title{The XMM deep survey in the CDF-S. IX. \\
An X-ray outflow in a luminous obscured quasar at $\mathbf{z\approx1.6}$}

\author{
C. Vignali\inst{1,2}
\and
K. Iwasawa\inst{3}
\and
A. Comastri\inst{2}
\and
R. Gilli\inst{2}
\and
G. Lanzuisi\inst{2,1}
\and
P. Ranalli\inst{4}
\and
N. Cappelluti\inst{2}
\and
V. Mainieri\inst{5,6}
\and
I. Georgantopoulos\inst{4}
\and
F.J. Carrera\inst{7}
\and 
J. Fritz\inst{8,9}
\and
M. Brusa\inst{1,2}
\and 
W.N. Brandt\inst{10,11,12}
\and
F.E. Bauer\inst{13,14,15}
\and
F. Fiore\inst{16}
\and
F. Tombesi\inst{17,18}
}

\institute{
Dipartimento di Fisica e Astronomia, Alma Mater Studiorum, Universit\`a degli Studi di Bologna, 
Viale Berti Pichat 6/2, 40127 Bologna, Italy
\email{cristian.vignali@unibo.it}
\and
INAF -- Osservatorio Astronomico di Bologna, Via Ranzani 1, 40127 Bologna, 
Italy 
\and
ICREA and Institut de Ci\`encies del Cosmos (ICC), Universitat de Barcelona 
(IEEC-UB), Mart\'i i Franqu\`es 1, 08028, Barcelona, Spain
\and 
Institute for Astronomy \& Astrophysics, Space Applications \& Remote Sensing, 
National Observatory of Athens, Palaia Penteli 15236, Athens, Greece
\and
European Southern Observatory, Karl-Schwarzschild-Straße 2, 85748, Garching, Germany
\and
Excellence Cluster Universe, Technische Universit\"{a}t M\"{u}nchen, Boltzmannstrasse 2, D-85748, Garching, Germany
\and
Instituto de F\'\i sica de Cantabria (CSIC-UC), Avenida de los Castros, 39005 Santander, Spain 
\and
Sterrenkundig Observatorium, Universiteit Gent, Krijgslaan 281, S9 9000, Gent, Belgium
\and
%Centro de Radioastronom\'\i a y Astrof\'\i sica, CRyA, UNAM, Campus Morelia, A.P. 3-72, C.P. 58089, Michoac\'an, Mexico
Instituto de Radioastronom\'\i a y Astrof\'\i sica, IRAf, UNAM, Campus Morelia, A.P. 3-72, C.P. 58089, Mexico
\and
Department of Astronomy \& Astrophysics, The Pennsylvania State University, University Park, PA, 16802, USA
\and
Institute for Gravitation and the Cosmos, The Pennsylvania State University, University Park, PA 16802, USA
\and
Department of Physics, Eberly College of Science, The Pennsylvania State University, University Park, PA, 16802, USA 
\and
Instituto de Astrof\'{\i}sica, Facultad de F\'{\i}sica, Pontificia Universidad Cat\'{o}lica de Chile, 306, Santiago 22, Chile
\and
Millennium Institute of Astrophysics, MAS, Nuncio Monse\~{n}or S\'{o}tero Sanz 100, Providencia, Santiago de Chile
\and
Space Science Institute, 4750 Walnut Street, Suite 205, Boulder, Colorado 80301
\and
INAF -- Osservatorio Astronomico di Roma, Via Frascati 33, 00040 Monteporzio 
Catone, Roma, Italy
\and
Department of Astronomy, University of Maryland, College Park, MD 20742-2421, USA
\and
NASA Goddard Space Flight Center, Greenbelt, MD, USA
}

\date{Received 2 Feb. 2015 / Accepted 21 Aug. 2015}

% \abstract{}{}{}{}{} 
% 5 {} token are mandatory

\abstract{In active galactic nuclei (AGN)-galaxy co-evolution models, AGN 
winds and outflows are often invoked to explain why super-massive black holes 
and galaxies stop growing efficiently at a certain phase of their lives. 
They are commonly referred to as the leading actors of feedback processes. 
Evidence of ultra-fast ($v\gtrsim0.05c$) outflows in the innermost regions of 
AGN has been collected in the past decade by sensitive \xray\ observations 
for sizable samples of AGN, mostly at low redshift. 
Here we present ultra-deep \xmm\ and \chandra\ spectral data of an obscured  
(\Nh$\approx2\times10^{23}$~cm$^{-2}$), intrinsically luminous  
(\lx$\approx4\times10^{44}$~erg~s$^{-1}$) quasar (named PID352) at 
$z\approx1.6$ (derived from the \xray\ spectral analysis) in the 
\chandra\ Deep Field-South. The source is characterized by an iron emission 
and absorption line complex at observed energies of $E\approx2-3$~keV. 
While the emission line is interpreted as being due to neutral iron 
(consistent with the presence of cold absorption), the absorption feature is 
due to highly ionized iron transitions (FeXXV, FeXXVI) with an outflowing 
velocity of $0.14^{+0.02}_{-0.06}$c, as derived from photoionization models.  
The mass outflow rate --  $\sim 2$~\msun~yr$^{-1}$ -- is similar to the 
source accretion rate, and the derived mechanical energy rate is 
$\sim9.5\times10^{44}$~\lum, corresponding to 9\% of the source bolometric 
luminosity. PID352 represents one of the few cases where indications of 
\xray\ outflowing gas have been observed at high redshift thus far. 
This wind is powerful enough to provide feedback on the host galaxy.}


\authorrunning{C. Vignali et al.}
\titlerunning{A high-velocity X-ray outflow at high redshift}

\maketitle

\section{Introduction}
\label{introduction}
According to some AGN/galaxy co-evolution models, along the cosmic history of galaxies there is a dust-enshrouded phase associated with rapid SMBH growth and active star formation, largely triggered by galaxy mergers and encounters (e.g., \citealt{silk_rees1998, dimatteo2005, menci2008, hopkins2008, zubovas_king2012, lamastra2013}). At the end of this obscured phase, massive quasar-driven outflows blow away most of the cold gas reservoir, creating a population of ``red-and-dead" gas-poor elliptical galaxies (e.g., \citealt{cattaneo2009}). Support for this picture -- at least for the most luminous AGN -- comes from observations of kpc-scale outflows of molecular gas (e.g., \citealt{feruglio2010, sturm2011, cicone2014}), as well as of neutral and ionized gas (e.g., \citealt{nesvadba2008, alexander2010, harrison2012, harrison2014, genzel2014, perna2015, brusa2015}). On smaller scales, ultra-fast outflows (UFOs, with velocities typically up to 0.1--0.4c) have been clearly detected in X-rays in a sizable sample of AGN at low redshift (e.g., \citealt{reeves2003, pounds2003, tombesi2010, tombesi2012a, tombesi2014, tombesi2015, giustini2011, patrick2012, gofford2013, gofford2015, king_a_2014, nardini2015, ballo2015}; see also \citealt{cappi2006} for an early review, and \citealt{fabian2012} for a more recent one) and also in a limited number of high-redshift quasars (e.g., \citealt{chartas2002, chartas2007, chartas2009, chartas2014, saez2009, lanzuisi2012}). 
Although original claims of blueshifted absorption features were strongly debated (see \citealt{vaughan_uttley2008}), recent high photon statistics in \xray\ spectra of both radio-quiet (e.g., \citealt{tombesi2010, tombesi2011, tombesi2012a, gofford2013}) and radio-loud AGN (e.g., \citealt{gofford2013, tombesi2014}) have undoubtedly shown that these outflows are not rare events at \xray\ wavelengths, with more than 40\% to 50\% of the investigated Seyfert galaxies in the local Universe being associated with such highly energetic phenomena (e.g., \citealt{tombesi2010, gofford2013}). According to models (e.g., \citealt{zubovas_nayakshin2014}), fast ionized winds may lose most of their kinetic energy after shocking the interstellar medium (ISM), thus cooling efficiently and transferring their ram pressure (hence momentum flux) to the ISM. In this context, large-scale outflows would be produced by the accelerated swept-up gas (e.g., \citealt{king2010b, faucher-giguere_quataert2012}). 

In a study based on the \xray\ spectral analysis of a sample of Type~1 Seyfert galaxies, \cite{tombesi2013} have recently suggested that \xray\ warm absorbers (WA) and UFOs are two aspects of the same fundamental physical process, which is related to wind acceleration likely occurring in the accretion disk (see also \citealt{fukumura2014}). In particular, there is a sort of ``continuity" in the physical properties of the high- (UFO) and low-velocity (WA) winds, with the former being produced in the inner regions of the accretion disk, characterized by higher ionization and column density, and the latter associated either with the outer parts of the accretion disk or with the AGN torus. While WAs have typical velocities below $\sim1~000$~\kms\ and low kinetic power (e.g., \citealt{blustin2005, mckernan2007}, but see also \citealt{crenshaw_kraemer2012}), UFOs are characterized by velocities $>0.05c$ up to $\sim$0.6c (this high value is observed in, e.g., APM~08279$+$5255, \citealt{saez2009, chartas2009}, and HS~1700$+$6416, \citealt{lanzuisi2012}) and considerably higher mechanical energy, so they possibly exert a significant impact on the host galaxy (e.g., \citealt{hopkins_elvis2010, king2010b, gaspari2011, zubovas_nayakshin2014}). Regardless of the exact launching site for the wind, the observed outflow properties require an efficient process of acceleration, such as radiation pressure through Compton scattering and, more importantly,  magneto-hydrodynamical processes (e.g., \citealt{king_pounds2003, proga_kallman2004, sim2008, oshuga2009, fukumura2010}; see also \citealt{higginbottom2014}). 

From the innermost regions of AGN (probed by X-rays) to the outer galactic scales (probed by observations in molecular lines), all of these outflows may provide significant feedback into the quasar host galaxy (e.g., \citealt{king2010a, king2011}; see also \citealt{pounds2014} for a recent review) and may be responsible for  quenching star formation (e.g., \citealt{canodiaz2012, cresci2015}) and, if the momentum is maintained through the shocks, for setting up the local black hole mass vs. bulge relation (e.g., \citealt{king2010b, zubovas_nayakshin2014, costa2014}).
However, despite the increasing observational evidence of outflows at different scales, the link between UFOs and large-scale outflows is today basically unknown, especially at high redshift. In this regard, X-ray emission offers a powerful probe of the innermost regions of the AGN where fast outflows are launched (e.g., \citealt{proga_kallman2004}). 

In this paper we present the intriguing properties of the source XMMCDFSJ033242.4$-$273815, 
whose \xmm\ and \chandra\ spectral data are consistent with the presence of an emission-plus-absorption line, although at a different statistical significance (lower in \chandra). The emission line is interpreted as neutral iron K$\alpha$ emission, and the absorption line is most likely due to a blueshifted highly ionized iron transition associated with outflowing gas with a velocity of $\sim$0.1c. 
The spectral ``peculiarity" of this source came out in the search for obscured AGN at high redshift selected in the \xmm\ observations of the \chandra\ Deep Field-South (XMM-CDFS) by means of restframe \xray\ colors \citep{iwasawa2012}. 
This source is listed as number 352 in the 3Ms \xmm\ 2--10~keV source catalog by \cite{ranalli2013}; hereafter, we refer to this source as PID352. 
In the following, we present the analysis of these iron features based on the whole dataset of \xmm\ and \chandra\ observations, carried out over a time interval of $\sim$10 years. 
Most of the spectral results presented in this paper rely on \xmm\ data, while \chandra\ provides further support to these results, as well as an accurate source position.

Currently, there are only two photometric redshift estimates for PID352, one from \cite{taylor2009}, $z_{\rm ph}=1.51^{+0.34}_{-0.20}$ (95\% confidence level), and the other from \cite{hsu2014}, $z_{\rm ph}=1.31^{+0.78}_{-0.74}$ (99\% confidence level).

\section{X-ray data}
\label{xray_data}
The CDF-S was observed with \xmm\ with 33 exposures over the years 2001--2002 and 2008--2010, for a nominal exposure time of $\sim$3.45Ms. Full details on the pointing strategy, data cleaning, and derived catalogs are presented by \cite{ranalli2013}. In the analysis presented in this work, we use data taken from the three EPIC cameras (pn, MOS1, and MOS2) after applying a sigma-clipping procedure to the event files in order to filter high-background flaring intervals. Given the large off-axis angle of PID352 in the summed \xmm\ mosaic,\footnote{The exact off-axis angle in the final \xmm\ mosaic is difficult to quantify, since the nominal right ascension and declination of the individual pointings varied substantially, with the highest difference being between the 2001--2002 observations and the most recent 2008--2010 observations.} the final effective (i.e., after filtering and accounting for vignetting effects) exposure time at the source position is $\sim$800~ks (averaged over the three cameras). 
The resulting 2--10~keV position of PID352 is (ra,dec)=(03:32:42.46,$-$27:38:15.7; 
\citealt{ranalli2013}).  
Spectral data were extracted from the three mosaics of pn, MOS1 and MOS2 using circular regions of 10\arcsec\ radius (corresponding to $\simlt$60\% of the encircled energy fraction  at the source position) in order to maximize the signal-to-noise ratio in the extraction regions. Background was chosen from nearby, source-free circular regions of radius  19\arcsec, 25\arcsec\,, and 25\arcsec\ for the pn, MOS1, and MOS2, respectively.  
Once normalized to the source extraction regions, the background accounts for $\sim$30--35\% of the overall counts. The final number of net (background-subtracted) source counts in the $\sim$0.3--7~keV band is 2560, 690, and 1250 in the pn, MOS1, and MOS2 cameras, respectively. 

PID352 was also observed in both of the \chandra\ CDF-S and Extended-CDF-S (E-CDF-S) exposures. CDF-S data were taken in the periods 1999--2000, 2007, and 2010 with 54 pointings using the ACIS-I array for a total on-axis exposure time of $\sim$4Ms; see \cite{luo2008} and \cite{xue2011} for more details on the observational strategy and data reduction. PID352 is associated with source XID~571 in the \cite{xue2011} catalog [with coordinates (ra,dec)=(03:32:42.73,$-$27:38:15.6)]. As a consequence of the large off-axis angle of the source ($>$10\arcmin) and the different roll angles of the CDF-S exposures, the source falls in the CDF-S in only 13 out of the total 54 \chandra\ pointings. 
The count distribution seems to be slightly displaced (by $\sim$1.1\arcsec) with respect to the \xray\ source position in the catalog, probably because of the off-axis position (hence elongated shape) of PID352 in the 4Ms exposure. Once this offset is taken into account, the source is perfectly coincident with a red galaxy (see $\S$\ref{multi}). 
The final vignetting-corrected source exposure time is $\sim$430ks.  
The \chandra\ spectrum was extracted from individual pointings using the {\sc Acis Extract} software \citep{broos2010} and accounting for the different point spread function (PSF) size and shape at the source position in each observation. 
The total number of net source counts in the $\sim$0.5--7~keV band is $\sim$450, which is fully consistent with the number reported in the \cite{xue2011} catalog. The overall contribution of the background (chosen from nearby source-free regions) is limited to less than 10\% of the source plus background counts within the source extraction region. 
We note that at $\sim$8.7\arcsec\ from its position there is another fainter (by a factor of $\sim3.5$ in the 0.5--8~keV band) \xray\ source, listed as XID~568 in \cite{xue2011}. The good agreement in terms of \xray\ spectral results from \xmm\ and \chandra\ data suggests that the contamination by XID~568 to PID352 in \xmm\ data is most likely marginal. 

PID352 was also detected in the E-CDF-S mosaic (four pointings with a nominal exposure of $\sim$250ks each; observations were taken in 2004) and was reported as source 437 in the \cite{lehmer2005} catalog. At the source position (ra,dec)=(03:32:42.63,$-$27:38:16.1), which corresponds to an off-axis angle of 5.9\arcmin, the effective exposure time is $\sim$190ks. The source spectrum was extracted from a circular region with a radius of $\sim$4.5\arcsec, while the background spectrum was extracted from a circular source-free region with radius 19.5\arcsec. The impact of the background is 4\% in the source extraction region. The total number of source net counts in the $\sim$0.5--7~keV band is $\sim$260, so it is fully consistent with the number reported in the \cite{lehmer2005}  catalog. 

In the following analyses, \xmm\ and \chandra\ (both CDF-S and E-CDF-S) results will be presented separately. Overall, \xmm\ data provide good counting statistics, while \chandra\ data offer good source positioning, low background contamination, and independent (though statistically limited) support for \xmm\ spectral results. 

X-ray spectral analysis is carried out using {\sc xspec} vs. 12.6.1 \citep{arnaud1996}. 
Errors are quoted at the 90\% confidence level for one interesting parameter (i.e., either $\Delta\chi^{2}$=2.71 or $\Delta$C=2.71; \citealt{avni1976, cash1979}, depending on the adopted statistic). All spectral fits include absorption due to the line-of-sight Galactic column density of N$_{\rm H}=7\times10^{19}$~cm$^{-2}$ \citep{kalberla2005},  and \cite{anders_grevesse1989} abundances are assumed. 
Hereafter we adopt a cosmology with $H_{0}$=70~km~s$^{-1}$~Mpc$^{-1}$, $\Omega_{\Lambda}$=0.7, and $\Omega_{M}$=0.3.

\subsection{XMM-Newton spectral results}
\label{xmmcdfs}
The \xmm\ spectra were binned at a signal-to-noise ratio of at least 3. We have verified that such binning satisfies the criterion of the minimum number of photons/bin typically adopted to apply $\chi^{2}$ statistics (e.g., 20--30 counts per bin). We have also checked that the results reported hereafter are recovered (within the errors) once a ``standard" binning is adopted. 
In the following, we adopt ``phenomenological" models to reproduce the \xmm\ data, then 
in $\S$\ref{discussion} we present results obtained with a more self-consistent and physically-motivated model.  

\begin{figure}
\centering
\includegraphics[angle=0,width=0.48\textwidth]{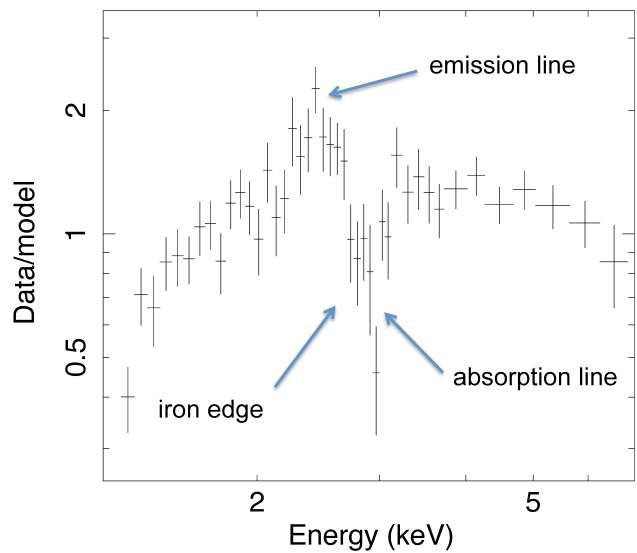}
\caption{Plot of the observed 1.3--7~keV band spectrum of PID352 divided by the best-fitting powerlaw model (with a flat photon index, see text for details). 
The spectral data were made by  combining the \xmm\ EPIC pn, MOS1, MOS2, and \chandra\ ACIS-I data. The powerlaw fit was performed to the data from all these instruments jointly. 
The most relevant spectral features are indicated in the figure.} 
\label{residuals_xmm}
\end{figure}

Fitting the \xmm\ data with a single powerlaw and Galactic absorption produces statistically significant data-to-model residuals ($\chi^{2}$/d.o.f=484.5/276, where d.o.f. is the number of degrees of freedom). 
From the spectral deviations shown in Fig.~\ref{residuals_xmm} (where the assumed model is a powerlaw), it is clear that additional spectral complexities are present at observed energies of $\sim$2--3~keV, namely an emission and an absorption feature, and an absorption edge.  The notably flat photon index, $\Gamma=0.45\pm{0.05}$, is suggestive of obscuration, and an additional component (parameterized by a second powerlaw) is required in the soft band. 
% ---------------------------------------------------------------------------
% ---------------------------------------------------------------------------

Thus we adopted a double powerlaw model with free photon indices; one of these components is absorbed by cold matter. The resulting fit provides $\chi^{2}$/d.o.f=345.7/273. 
The two photon indices are consistent (within $\sim1.1\sigma$) considering their errors ($\Gamma_{\rm soft}=1.94^{+0.36}_{-0.34}$ and $\Gamma_{\rm hard}=1.53^{+0.22}_{-0.20}$), while the column density ($pha$ model in {\sc xspec}) at $z$=0 is N$_{\rm H}=(1.53^{+0.37}_{-0.31})\times10^{22}$~cm$^{-2}$. 
If the soft \xray\ emission is interpreted as scattering, this modeling results in a $\sim$6\% of scattering fraction, which is a bit high but possible based on the observed range in \xray\ spectra of obscured AGN (e.g., \citealt{lanzuisi2015}).
Since we are mostly interested in the source emission and absorption features and the hard \xray\ emission, we do not investigate this issue further. 
% ---------------------------------------------------------------------------
\begin{figure*}
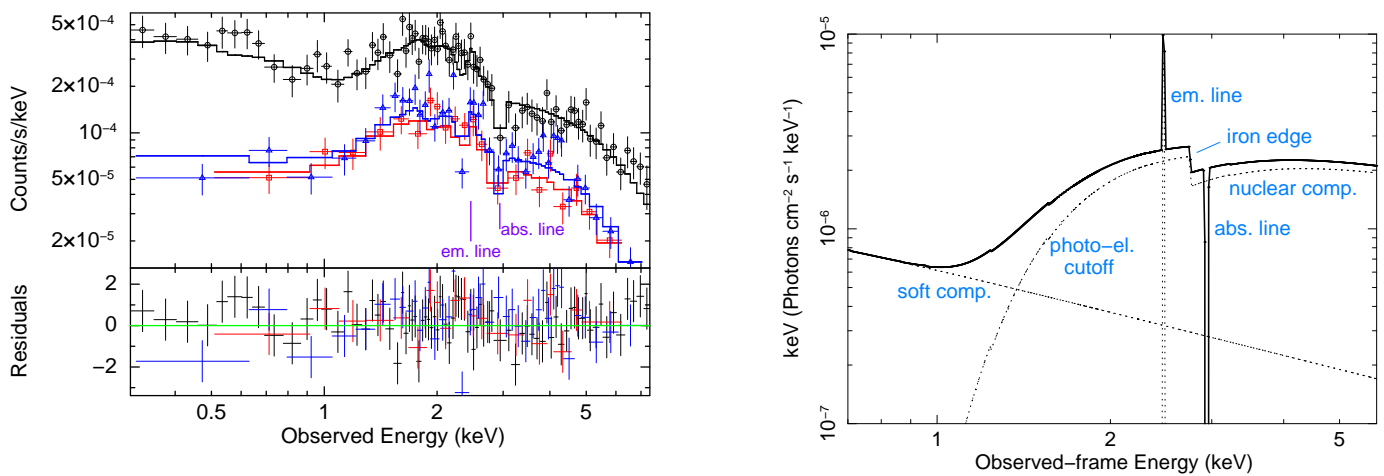

\centering
\includegraphics[angle=-90,width=0.48\textwidth]{vignali_fig2a.eps}\hfill
\includegraphics[angle=-90,width=0.45\textwidth]{vignali_fig2b.eps} \caption{{\em (Left panel)} \xmm\ spectrum of PID352 fitted with a double powerlaw (one absorbed) plus one emission line and one absorption line. The black/blue/red lines and datapoints refer to pn/MOS1/MOS2 data. Emission and absorption lines are both marked. In the bottom panel, the data-to-model residuals are shown in units of $\sigma$. {\em (Right panel)} Best-fitting \xmm\ model (thick solid line), with indications of the spectral components (dotted lines). In particular, we note that the iron edge is at a different energy with respect to that of the absorption feature.}
\label{spec_xmm}
\end{figure*}
% ---------------------------------------------------------------------------
%
% ---------------------------------------------------------------------------
\begin{figure*}
\centering
\includegraphics[angle=-90,width=0.48\textwidth]{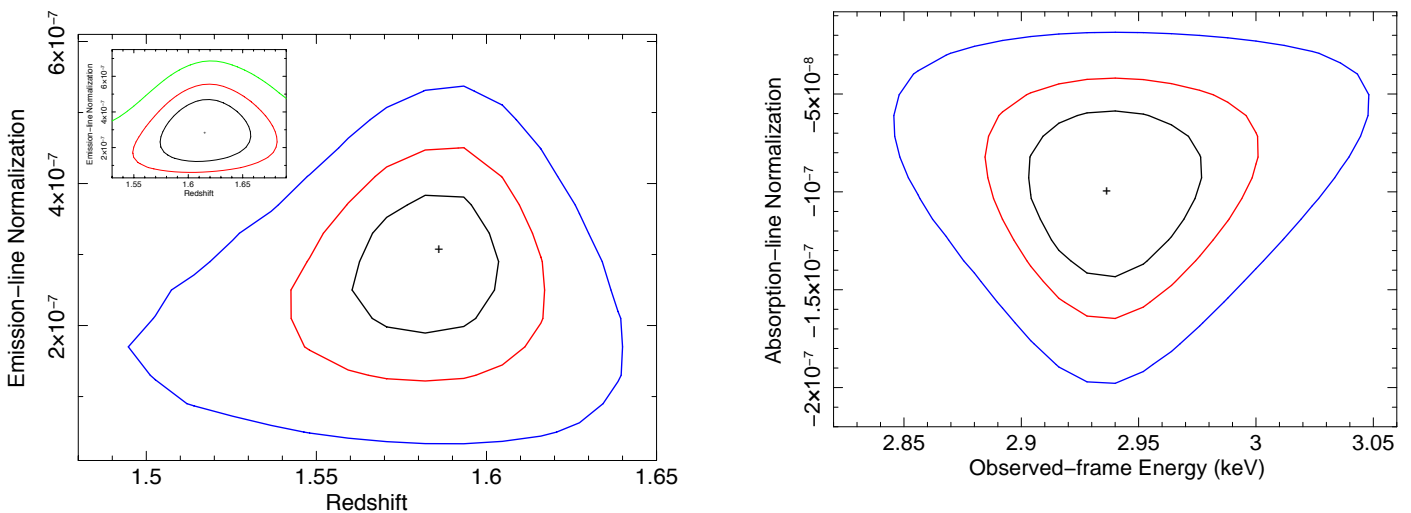}\hfill
\includegraphics[angle=-90,width=0.48\textwidth]{vignali_fig3b.eps}
\caption{\xmm\ spectral results: {\em (left panel)} redshift vs. FeK$\alpha$ line normalization. For comparison, the redshift estimate obtained by \chandra\ data is shown in the inset; {\em (right panel)} observed-frame absorption-line energy vs. normalization. In all panels, contours represent the 68, 90, and 99\% confidence levels for two parameters of interest.}
\label{lines_xmm}
\end{figure*}
% ---------------------------------------------------------------------------
%\input{vignali_table1.tex}
%----------------------------------------------------------------------------
\begin{table*}
\small
\caption[]{Best-fitting \xray\ spectral parameters of PID352 derived from \xmm\ and \chandra\ (CDF-S and E-CDF-S).}
\label{xray_results}
%\resizebox{0.485\textwidth}{!}{                                                                                                                                 
\begin{tabular}{lccccccccccc}
\hline
\noalign{\smallskip}
\small
Data & $\Gamma_{\rm soft}$ & $\Gamma_{\rm hard}$ & N$_{\rm H}$  & $z$ & EW$_{\rm em. line}$ & E$_{\rm abs. line}$ & EW$_{\rm abs. line}$ & F$_{2-10~keV}$ & L$_{\
2-10~keV}$ & $\chi^{2}$(C)/d.o.f.\\
         &          &            & ($\times10^{23}$cm$^{-2}$) &     & (eV)              & (keV)             & (eV)             & (\cgs)        & (\lum)       & \
\\
\hline
\noalign{\smallskip}
{\bf X} & 1.73$\pm{0.31}$ & 1.58$^{+0.28}_{-0.23}$ & 2.30$^{+0.76}_{-0.54}$ &
1.59$\pm{0.03}$ & 76$\pm{35}$ & 2.94$\pm{0.04}$ & $-$82$^{+36}_{-39}$ & 2.6--3.1--2.8 & 3.5--4.2--3.8 & 275.3/268 \\
{\bf C} & ... & 1.03$^{+0.29}_{-0.26}$ & 1.01$^{+0.42}_{-0.34}$ &
1.62$\pm{0.05}$ & 120$^{+83}_{-73}$ & 2.85$^{+0.16}_{-0.07}$ & $-$65$^{+61}_{-48}$ & 2.4--3.6 & 1.9--2.8 & 350.0/480 \\
\noalign{\smallskip}
\hline
\end{tabular}
\tablefoot{X: \xmm\ (pn, MOS1, and MOS2); C: \chandra\ (CDF-S 4Ms$+$E-CDF-S data). Both line energies and equivalent widths
are reported in the observed frame, while the column density is given at
the reported best-fitting redshift solution.
%Fluxes (luminosities) are reported in units of 10$^{-14}$~\cgs\ and 10$^{44}$~\lum, respectively.                                                               
Fluxes are reported in the observed 2--10~keV band (units of 10$^{-14}$), while luminosities are intrinsic (i.e., corrected for obscuration) and in the restframe 2--10~keV band (units of 10$^{44}$). The three fluxes and luminosities reported for \xmm\ refer to pn, MOS1, and MOS2 data, respectively, while for \chandra\ \
the two fluxes and luminosities refer to CDF-S and E-CDF-S data, respectively.
Both emission and absorption lines were modeled using a Gaussian feature of width $\sigma$=10 eV.}
\end{table*}
%----------------------------------------------------------------------------
%
Then we modified the previous model by leaving the redshift of the absorber free to vary.  We obtain $z=1.59^{+0.06}_{-0.07}$ and powerlaw slopes consistent with the values reported above within their errors;  the derived column density is N$_{\rm H}=(2.87^{+0.97}_{-0.69})\times10^{23}$~cm$^{-2}$. The resulting fit quality is $\chi^{2}$/d.o.f=301.7/272. 
The emission and absorption features seem to be too separated in energy to be a P-Cygni profile. 
Therefore, to reproduce these features, we start by adding a narrow ($\sigma=10$~eV) emission line to the previous modeling, leaving the redshift as a free parameter. Constraints on redshift come from the emission line (assumed to be associated with the neutral restframe 6.4~keV iron transition, which is expected given the presence of cold absorption) and the iron K$\alpha$ edge (at the restframe energy of 7.1~keV, seen at an observed energy of $\sim$~2.7~keV in Fig.~\ref{residuals_xmm}) associated with the absorber. Including the line (observed equivalent width $EW_{\rm obs}=76\pm{35}$~eV, corresponding to $EW_{\rm rest}\approx$200~eV at the derived redshift of $z=1.59\pm{0.03}$) improves the quality of the spectral fit ($\chi^{2}$/d.o.f=289.3/271, i.e., $\Delta\chi^{2}$/$\Delta$d.o.f=12.4/1 with respect to the previous double powerlaw plus absorption model). The line EW appears only marginally consistent (within errors) with the EW expected in the case of transmission through a neutral shell of gas according to the recent Monte Carlo calculations provided by \cite{murphy_yaqoob2009}, 
but seems to agree more with past-work predictions (e.g., \citealt{awaki1991,  leahy_creighton1993}). 

The \xray\ spectral slopes do not change significantly with respect to the previous modeling once errors are taken into account. The column density at the newly derived redshift is N$_{\rm H}=(2.30^{+0.76}_{-0.54})\times10^{23}$~cm$^{-2}$.

The following step consists of modeling the absorption feature with a narrow (10~eV) Gaussian line. This provides a spectral improvement of $\Delta\chi^{2}$/$\Delta$d.o.f=14/2  ($\chi^{2}$/d.o.f=275.3/269). 
Monte Carlo simulations indicate that the probability of by chance detecting an absorption line providing an improvement to the spectral fitting larger or comparable to the one we obtain is $\sim$1\% (see $\S$\ref{xray_spectral} for details). 
Most of the parameters related to the \xray\ spectrum remain unchanged. The observed-frame  energy and EW of the absorption line are E=$2.94\pm{0.04}$~keV and $-82^{+36}_{-39}$~eV, respectively. 

The best-fitting \xmm\ pn, MOS1, and MOS2 spectra are shown in Fig.~\ref{spec_xmm} (left panel). The redshift solution is shown in Fig.~\ref{lines_xmm} (left panel; $z=1.59^{+0.03}_{-0.05}$ according to 90\% confidence contours), while the observed-frame energy vs. normalization of the absorption line is shown in Fig.~\ref{lines_xmm} (right panel). The confidence contours indicate that both the source redshift and the absorption feature are well constrained in the \xmm\ analysis, which is considerably better than using \chandra\ data (although consistent in terms of line parameters, see $\S$\ref{cdfs}). 
It is worth mentioning that the energy of the neutral iron absorption edge (E=2.74~keV) differs clearly from that of the absorption line (see Fig.~\ref{spec_xmm}, right panel). 

At face value, the derived blueshift of the absorption line (most likely ascribed to ionized iron resonant absorption), assuming the redshift of $z=1.59$, is $\sim$0.13c and $\sim$0.09c in case of He-like (FeXXV, 6.7~keV) and H-like (FeXXVI, 6.97~keV) iron transitions, respectively. The association with either He- or H-like iron transitions is the most likely, according to previous, higher statistics studies of AGN showing outflows (e.g., \citealt{risaliti2005}); see $\S$\ref{discussion} for a complete discussion on this issue using a more physically motivated modeling of \xmm\ data. 

We have also verified whether different modeling could account for the source complex emission; in particular, motivated by the observed flat photon index (which may be partly related also to the presence of radio emission, see $\S$\ref{radio}), we checked for a reflection component by using the {\sc pexmon} model within {\sc xspec} \citep{nandra2007}. This model includes Compton reflection with self-consistently generated fluorescence emission lines from iron and nickel plus the FeK$\alpha$ Compton shoulder. 
We set the {\sc pexmon} model to only produce the reflection component by making the relative reflection parameter $R$ negative and by tying its photon index and normalization to the corresponding parameters of the primary powerlaw. We fixed the high-energy cutoff of the illuminating powerlaw at 100 keV, the abundances of the distant reflector to the solar values, and the inclination angle to 60 degrees. This model does not provide a good fit to the data ($\chi^{2}$/d.o.f=323.8/270), because the reflection/powerlaw normalization ratio is unphysically high and basically unconstrained, as the entire spectrum of PID352 were reflection-dominated (which is also not consistent with the observed emission-line EW). Similar results have been obtained using {\sc pexrav} \citep{magdziarz_zdziarski1995} plus an emission line. 

The observed-frame 2--10~keV source flux is $\sim(2.6-3.1-2.8)\times10^{-14}$~\cgs\ (where the three values refer to pn, MOS1, and MOS2, respectively, owing to their slightly different  normalizations). Once corrected for intrinsic absorption, the restframe 2--10~keV source luminosity is $\sim(3.5-4.2-3.8)\times10^{44}$~\lum.
The best-fitting \xmm\ results are summarized in Table~\ref{xray_results}.

\subsection{Chandra spectral results: CDF-S and E-CDF-S}
\label{cdfs}
We then checked for indications of the iron line complexity in PID352 using the available \chandra\ spectra, well aware that the limited photon statistics will prevent us from drawing 
firm conclusions on this issue from \chandra\ data alone.  
At first, the CDF-S and E-CDF-S spectral data of PID352 were fitted separately in order to appreciate,  from the admittedly low-counting statistics spectra, any possible significant difference, mostly in the continuum emission, which might preclude a joint analysis. A powerlaw model was adopted at this stage. The datasets provided consistent spectral results in terms of photon index (within errors), while the source flux is a factor $\sim$1.4 higher in the E-CDF-S data than in CDF-S data. In all of the following analyses, we proceed with simultaneous \xray\ spectral fitting of both datasets, leaving the normalizations free to vary and using the Cash statistic (i.e., using unbinned data; see \citealt{cash1979}). 

The flat photon index obtained adopting a powerlaw model, $\Gamma=0.2\pm{0.1}$, is suggestive of the presence of obscuration toward the source, similar to \xmm\ data analysis. To properly investigate this issue, we included an obscuring screen in the spectral
modeling, which returns a column density of (9.6$^{+3.5}_{-3.0})\times10^{21}$~cm$^{-2}$ at redshift $z$=0. As expected, the photon index becomes steeper ($\Gamma=1.2\pm{0.3}$), although not  as steep as typically found in unobscured AGN and quasars ($\Gamma\approx1.8$, e.g., \citealt{piconcelli2005}). The shape of the data-to-model residuals, especially in the higher statistics CDF-S spectrum, are suggestive of some spectral features at observed energies of $\sim2-3$~keV, although the relevance of such features is clearly limited by the paucity of photons. 

Then we included an emission line, whose best-fitting observed-frame energy is 2.44$^{+0.04}_{-0.05}$~keV, consistent with iron K$\alpha$ emission if the redshift derived from \xmm\ data analysis is assumed. The improvement with respect to the previous fit in terms of $\Delta$C is limited (10 for two less d.o.f., i.e., the line energy and its normalization, since the line width has been frozen to $\sigma$=10~eV). 

We note that other data-to-model residuals may reflect the presence of an absorption line, similar to what has been observed in \xmm\ data. Once modeled as a narrow (10~eV) Gaussian line, the fit improves slightly  ($\Delta$C/$\Delta$d.o.f.$\approx$5/2). The observed-frame energy of this additional line is 2.85$^{+0.16}_{-0.07}$~keV. 

\begin{figure}
\centering
\includegraphics[angle=-90,width=0.48\textwidth]{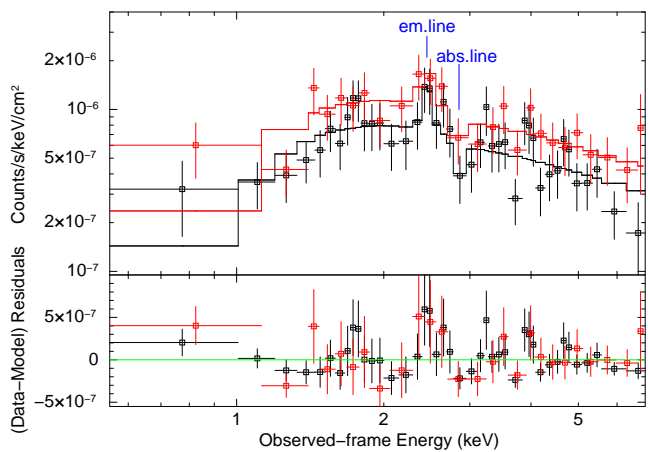}
\caption{{\em (Top panel)} \chandra\ CDF-S and E-CDF-S spectra of PID352, normalized by the effective area and fitted with an absorbed powerlaw, plus one emission line and one absorption line. The red datapoints refer to the E-CDF-S data, and the black ones to the higher statistics CDF-S data. The best-fitting model is shown as continuous curves. The observed-frame energy of the emission and absorption lines are also shown. All of the data have been rebinned to the 3$\sigma$ level for presentation purposes. 
{\em (Bottom panel)} Data $-$ model residuals once both datasets are fitted with an absorbed powerlaw without including any feature.}
\label{spec_chandra}
\end{figure}

As for the \xmm\ spectral analysis ($\S$\ref{xmmcdfs}), we then decided to model the data assuming that the iron K$\alpha$ line is neutral (6.4~keV restframe) and leaving its redshift free to vary. The redshift solution is therefore determined by the iron emission line and the iron absorption edge due to neutral matter. The derived redshift is $z=1.62\pm{0.05}$ ($z=1.62^{+0.08}_{-0.07}$ if one considers the 90\% confidence level contours plotted in the inset of Fig.~\ref{lines_xmm}, left panel). 
The column density at the newly determined source redshift is N$_{\rm H}=(1.01^{+0.42}_{-0.34})\times10^{23}$~cm$^{-2}$, while the photon index apparently remains flat ($\Gamma=1.0\pm{0.3}$), possibly suggesting additional absorption complexity. Assuming $\Gamma$=1.8, the column density becomes $\sim2\times10^{23}$~cm$^{-2}$. 
The observed-frame emission (absorption) line EW is 120~eV ($-$65) eV, with large uncertainties (see Table~\ref{xray_results}) owing to the low statistical significance of both lines. 
The best-fitting \xray\ spectrum (rebinned at the 3$\sigma$ level for presentation purposes) is shown in Fig.~\ref{spec_chandra}, where E-CDF-S data are shown in red and CDF-S data in black, both normalized by the effective area. The energy of both features is marked. At the redshift of $z=1.62$ and assuming that the absorption feature is due to either He-like or H-like iron transitions, we derive blueshift velocities of $\sim$0.11c and $\sim$0.07c, respectively. 

The flux of the source in the observed 2--10~keV band is  $\sim(2.4-3.6)\times10^{-14}$~\cgs, corresponding to a restframe 2--10~keV intrinsic (i.e., corrected for the measured absorption) luminosity  of  $\sim(1.9-2.8)\times10^{44}$~\lum\ (where the two values refer to CDF-S and E-CDF-S data, respectively). 
The best-fitting \chandra\ results are summarized in Table~\ref{xray_results}. 

% ---------------------------------------------------------------------------
% ---------------------------------------------------------------------------

\subsection{What \xmm\ and \chandra\ tell us about PID352}
\label{xray_spectral}
From the \xray\ spectral analyses reported above and summarized in Table~\ref{xray_results}, it is clear that the \xmm\ and \chandra\ results are consistent within errors, both in terms of continuum emission and emission/absorption features. The \xmm\ flux lies between that of CDF-S and E-CDF-S data. The only significant spectral difference is the presence of an additional soft component, parameterized by a powerlaw, in the \xmm\ data; it seems plausible that we are able to observe such a component in \xmm\ because of its higher effective area with respect to \chandra. The hard powerlaw is flatter in \chandra\ than in \xmm\ and may explain the apparently higher column density derived in the \xmm\ spectral analysis (see Table~\ref{xray_results}). Once $\Gamma=1.8$ is assumed, the column densities derived from \chandra\ and \xmm\ data become similar. We note that applying the best-fitting \xmm\ model to the \chandra\ data (allowing the normalization of the continuum to be different) provides an acceptable fit. 
Also the features associated with iron have similar properties in \xmm\ and \chandra\ spectra.

Assuming the best-fitting results from \xmm\ analysis (owing to the higher quality of the data) and $z=1.59\pm{0.03}$, we are able to derive the following outflow velocities for the gas: $v_{\rm out}=0.13\pm{0.02}$c and 0.09$^{+0.02}_{-0.03}$c in case of He-like (FeXXV, 6.7~keV) and H-like (FeXXVI, 6.97~keV) iron transitions, respectively. We postpone the discussion on the ionization state of the line to Sect.~\ref{discussion}. 

Although the absorption feature is observed in both \xmm\ and \chandra\ datasets, one may argue about its statistical significance. To this point, extensive Monte Carlo simulations were carried out by producing 10~000 spectra for each EPIC instrument (pn, MOS1,  and MOS2) using the {\sc fakeit} routine in {\sc xspec} (e.g., \citealt{lanzuisi2013}). For each camera, the simulated spectra have the same background flux and exposure as observed. 
As input model for our simulations, we assumed the best-fitting model obtained in Sect.~\ref{xmmcdfs} deprived of the absorption feature. Therefore, the significance of the absorption line is estimated by computing how many times this feature is recovered by chance  and produces an improvement in the spectral fitting larger or comparable to the one actually measured using real data. In practice, we first fitted the simulated spectra with the same model used for the simulation with free parameters and then introduced an absorption line. 
The line energy was left free to vary in the full 0.3--7~keV range and constrained to be the same in all three (pn, MOS1, and MOS2) spectra. To avoid the fitted energy of the absorption line clustering around the initial value, for each simulation we repeated the fit with an additional absorption line 68 times with initial energies in steps of 0.1~keV between 0.3 and 7 keV, and selected the best fit among them. Only 101 of the 10~000 simulated spectra show a spectral improvement larger than or equal to what is observed ($\chi^2\approx14$). This translates into a probability of detecting an absorption line when it does not exist of 1.01\%. 
We then used a Bayesian procedure to find the probability distribution of the fraction of spurious detections (see example in pages 20–-22 of \citealt{wall_jenkins2003}) from 101 spurious detections out of 10~000 trials. This probability distribution was used to estimate confidence intervals in $f$, by determining the narrowest interval around the mode $\hat{f}=0.0101$ (S. Andreon, private communication) that included the corresponding probability.  
We obtained $f<0.0132$ at 99.73\% confidence.

As \cite{vaughan_uttley2008} point out, relatively strong lines with large uncertainties (low EW/error ratio) may suffer from the so-called publication bias\footnote{The publication bias consists in the fact that only the observations with detected features have been reported in the literature. In other words, the reported detections of emission/absorption lines may be just the ``tip of the iceberg", i.e., the strongest or most significant ones from a distribution of random fluctuations.} (but see also \citealt{tombesi2010}). 
In the XMM-CDFS source catalog \citep{ranalli2013}, there are only ten sources with more than 5000 net counts, PID352 being exactly in the tenth position in terms of number of counts. \xray\ spectral analysis of this high-statistics sample (Comastri et al., in preparation; see \citealt{iwasawa2015} for a detailed study of the two brightest AGN) shows no indication of absorption features as the one detected in PID352. We can safely assume that below this number of counts, the detection of absorption features would be challenging. 
Considering the size of this \xmm\ highly reliable ``spectroscopic" sample and the discussion reported in $\S$4.4 of \cite{vaughan_uttley2008}, we can argue that we might expect 0.1 such detections above the $\sim$2.6$\sigma$ significance of the line. A comparable number is obtained if we consider also the four sources with more than 4000 net counts. The presence of such a feature in the different EPIC cameras and, possibly, in the \chandra\ data makes us confident of the reliability of the line, although its statistical significance is admittedly limited. 

Furthermore, we note that at the energy of the emission and absorption features, there are no calibration issues in the effective area of \xmm\ instruments. The accuracy is better than 3\% and 2\% in MOS and pn cameras, respectively (see XMM-SOC-CAL-TN-0018). The main response artifacts (i.e., the Au M and Ir K edges in the effective area) can therefore not be responsible for the observed features. 

% ---------------------------------------------------------------------------
\begin{figure*}
\centering
\includegraphics[angle=0,width=0.48\textwidth]{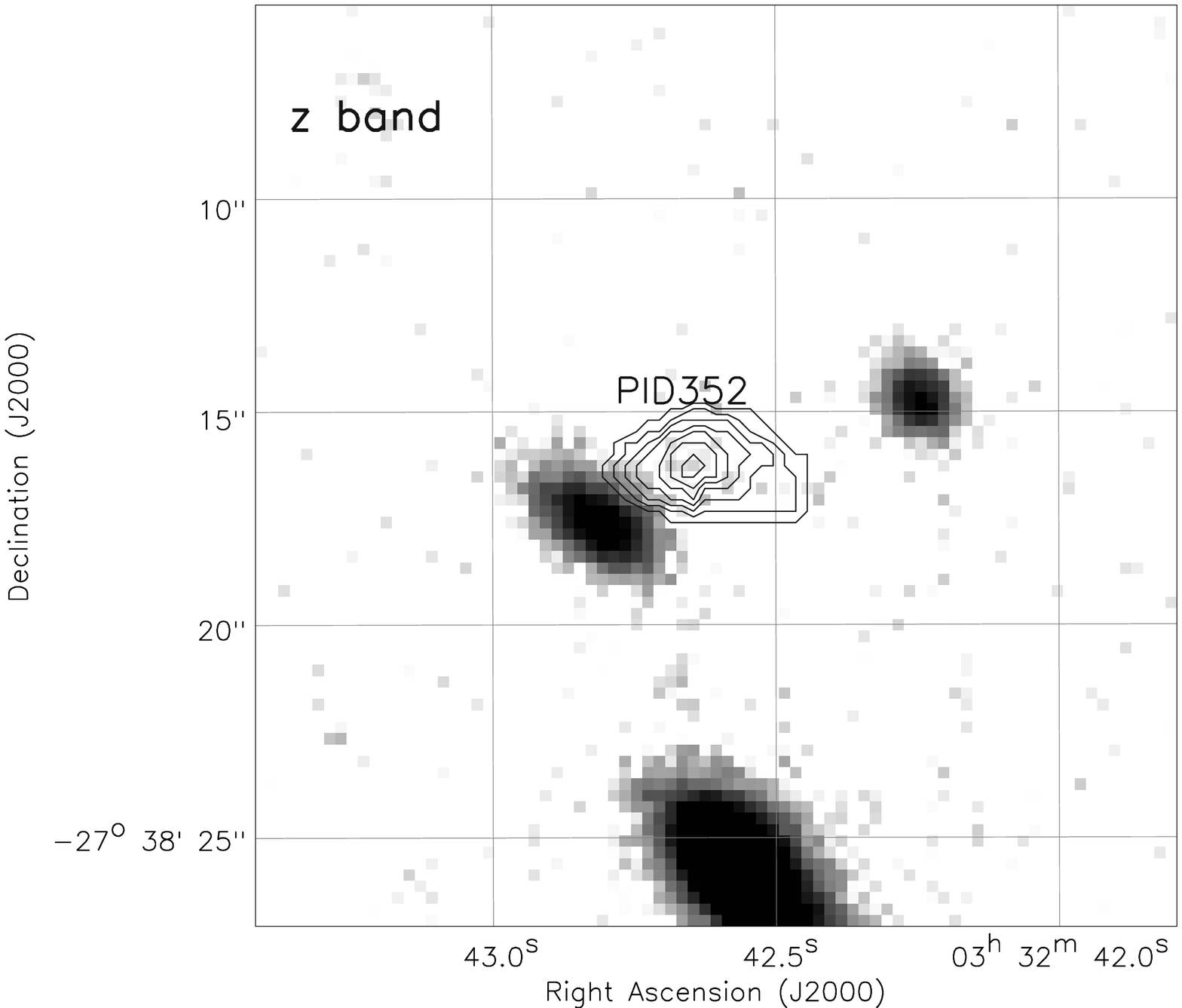}
\hfill
\includegraphics[angle=0,width=0.48\textwidth]{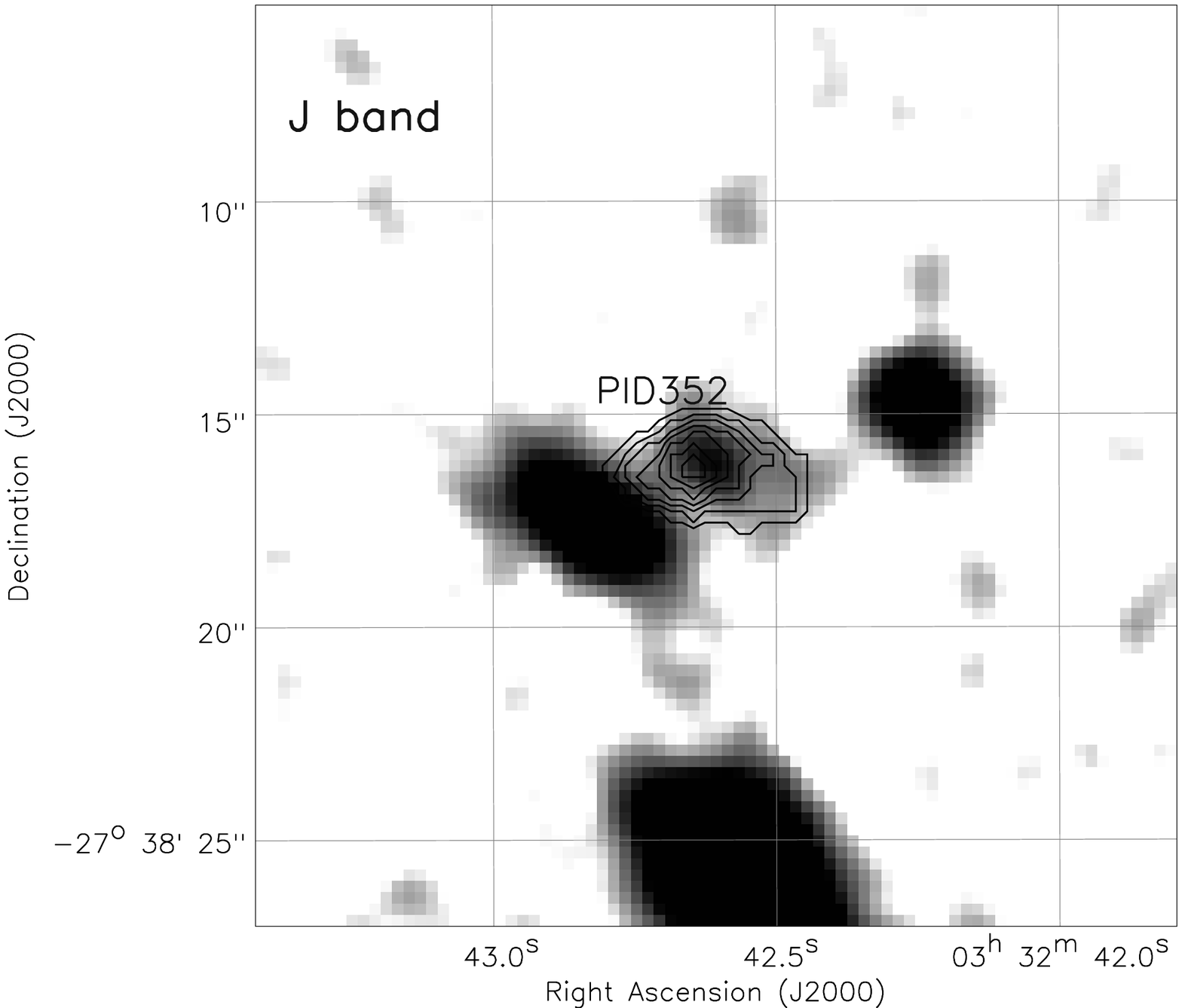}
\caption{MUSYC {\it z}-band (left panel) and {\it J}-band (right panel) smoothed images with the \chandra\ contours of source PID352 overlaid. The \xray\ source is associated with a red galaxy clearly appearing in filters redward of the {\it z} band. Each image is $\sim22$\arcsec\ by side. North is up, and east to the left.}
\label{musyc_xray}
\end{figure*}
% ---------------------------------------------------------------------------

\section{Multiwavelength data}
\label{multi}
\subsection{Spectral energy distribution of PID352}
\label{sed}
\begin{figure}
\centering
\includegraphics[angle=-90,width=0.48\textwidth]{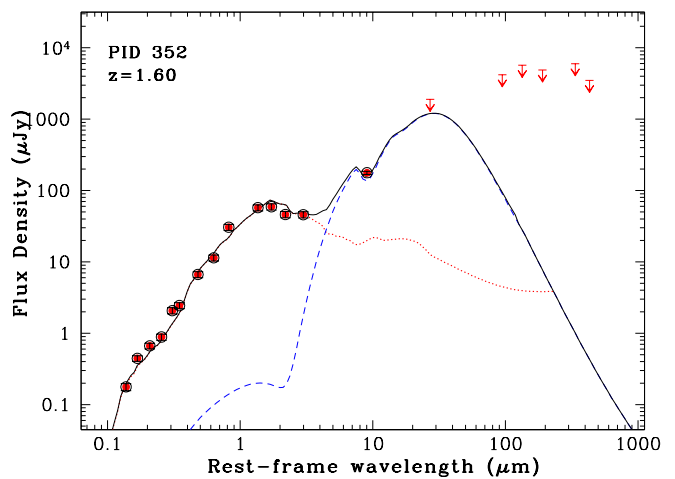}
\caption{Restframe spectral energy distribution for PID352. Available 
datapoints and upper limits (5$\sigma$) are plotted as red filled circles and downward-pointing arrows, respectively. The solid black line is the summed contribution of the AGN (blue dashed) and galaxy (red dotted) components. A redshift of $z=1.6$ has been assumed.}
\label{sed_agn}
\end{figure}

Using the available multiwavelength data for PID352 (from MUSYC, \citealt{cardamone2010}, \spitzer, \herschel\ SPIRE, \scuba, \laboca,\ and \aztec), we were able to characterize the source more properly. In particular, PID352 is associated with an extremely red object (ERO,  \citealt{elston1988}), because it has an {\it R$-$K} color (AB mag) of $\sim$3.8 (5.4 in Vega mag). This class of sources represents one of the major components of the stellar mass build-up at  redshifts $z\approx1-2$ (e.g., \citealt{daddi2000}). The source flux density in the observed MUSYC {\it J} band is a factor $\sim$3 higher than in the {\it z} band, where PID352 is barely visible (see Fig.~\ref{musyc_xray}; the source position is shown as \xray\ contours provided by \chandra\ CDF-S data). 
To have a more comprehensive picture of PID352 and derive an estimate of the source  bolometric luminosity, we modeled the source UV to millimeter emission in terms of galaxy plus AGN components (see Fig.~\ref{sed_agn}).

The spectral energy distribution (SED) of PID352 was modeled using the SED-fitting code originally developed by \cite{fritz2006} and recently updated by \cite{feltre2012}, and assuming $z=1.60$ as derived from our \xray\ spectral analysis. 
We note that, within the errors, the \xray-derived redshift is consistent with both current photometric-redshift estimates \citep{taylor2009, hsu2014}, and it is currently the one with the lowest uncertainties. 
The code accounts for both the stellar and the AGN components (for applications of this code, see \citealt{vignali2009, pozzi2012, gilli2014}). The stellar component is composed of a set of simple stellar population spectra of solar metallicity and ages up to $\sim$3~Gyr, i.e., the time elapsed between $z_{\rm form}$=6 (the redshift assumed here for the stars to form) and $z$=1.6 (the source redshift). This component is extincted according to a \cite{calzetti2000} attenuation law. 
For the AGN component, we used an extended grid of ``smooth" torus models with a ``flared disc"  geometry \citep{fritz2006}, which provides a good description of the AGN mid-IR SED in the case of sparse photometric datapoints (see also \citealt{vignali2011} and \citealt{feltre2012}). 

The galaxy provides the dominant contribution at short wavelengths (dotted line in Fig.~\ref{sed_agn}), where the best-fitting torus solution is suggestive of an obscured AGN (dashed line in Fig.~\ref{sed_agn}), i.e., an AGN providing a limited contribution to the optical emission because of extinction. The amount of obscuration toward the AGN derived from the SED fitting is fully consistent with the absorption found from \xray\ spectral analysis (assuming a Galactic dust-to-gas ratio conversion). However, the properties of the AGN component that may be derived from the SED fitting are not well constrained because of the lack of additional mid-IR data besides the \spitzer\ MIPS 24\micron\ datapoint. 
% ---------------------------------------------------------------------------
% ---------------------------------------------------------------------------

This said, we used the restframe 12.3\micron\ flux density and the \cite{gandhi2009} relation to estimate the predicted intrinsic 2--10~keV luminosity of the source. We obtained a value of $\sim2.5\times10^{44}$~\lum, which is $\sim$40\% lower than actually measured from \xmm\ spectral analysis but overall consistent if one considers the scatter in this mid-IR vs. \xray\ correlation. 

Although the source is apparently detected at 70\micron\ in the \herschel\ catalog \citep{magnelli2013}, the presence of a nearby far-IR bright source and the lack of any further detection of PID352 at longer wavelengths (including \laboca\ and \aztec\ data at 870\micron\ and 1100\micron) suggest to consider PID352 as undetected in the far-IR/mm (as shown in the SED plotted in Fig.~\ref{sed_agn}). 

\subsection{Radio data}
\label{radio}
\begin{figure}
\centering
\includegraphics[angle=0,width=0.48\textwidth]{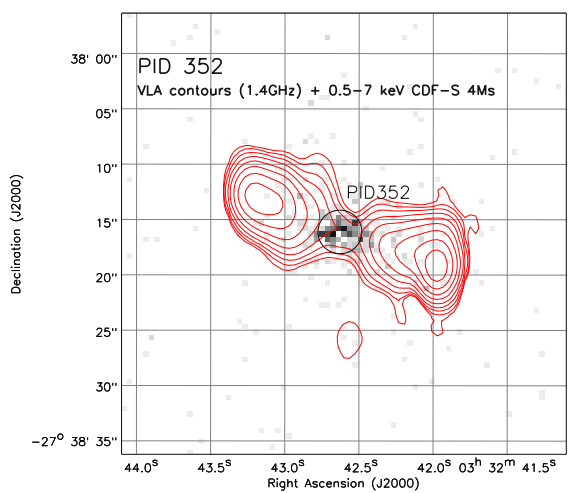}
\caption{CDF-S 0.5--7~keV unsmoothed image (40\arcsec\ by side) with the 20~cm contours (from \citealt{miller2013}) overlaid. \xray\ counts are most likely ascribed to the core of the radio galaxy, where no appreciable radio emission is detected (see \citealt{miller2013}). 
The \chandra\ position of PID352 (see $\S$\ref{xray_data} for details) is shown as a circle with a radius of 2\arcsec. North is up, and east to the left.}
\label{radio_x}
\end{figure}

PID352 has a long history of observations in the radio band. From the analysis of \vla\ 20cm (1.4GHz) and 6cm (4.8GHz) maps, \cite{miller2008} and \cite{kellermann2008} report the presence of an extended radio galaxy, classified as a double Fanaroff-Riley (FRII, \citealt{fanaroff_riley1974}), whose ``centroid"  corresponds to the red galaxy associated with PID352 (see Fig.~\ref{radio_x}). However, no radio emission is detected at the position of the core of this FRII galaxy, although in the 6cm map there are hints of emission at that position ($\sim0.64\pm{0.37}$~mJy). 
In the latest published version of the 20cm catalog by \cite{miller2013}, characterized by an average rms of $\sim7.4 \mu$Jy and a beam size FWHM$\approx2.8\arcsec\times1.6\arcsec$, the western and eastern radio lobes have  integrated flux densities of 46.1~mJy and 36.2~mJy, respectively (with errors of $\sim$30~\mjy). The radio power at 1.4GHz is thus $\sim9.3\times10^{26}$~W/Hz \citep{bonzini2012, bonzini2013}. 
If we adopt the measurements of the two radio lobes from \cite{kellermann2008} at 6cm, we derive an energy index\footnote{Here the flux density is reported as F$_{\nu}\propto\nu^{-\alpha}$.} of $\alpha\approx0.9-1.0$, which is consistent with the relatively old electron populations typically expected in radio lobes. 

The source is also reported in the ATLAS survey (carried out with \atca) by \cite{norris2006} and \cite{hales2014} at 20cm with a different resolution ($\sim11\arcsec\times15\arcsec$ and $\sim12\arcsec\times6\arcsec$, respectively). 
The flux densities reported in the two \atca-based papers for the western lobe are $44.9\pm{9.0}$~mJy and $55.4\pm{2.8}$~mJy, respectively, while those for eastern lobe are $27.4\pm{5.5}$~mJy and $38.8\pm{1.9}$~mJy, respectively. 
Not surprisingly, PID352 is undetected in the \vlba\ observations at 20cm by \cite{middelberg2011}. 

% ---------------------------------------------------------------------------
% ---------------------------------------------------------------------------

\section{Properties of the outflow: location, mass-outflow rate vs. accretion rate, mechanical energy, and efficiency} 
\label{discussion}
After having characterized the \xray\ data of PID352 using a ``phenomenological" model, we modeled the \xmm\ data using grids from the photoionization code {\sc xstar} \citep{bautista_kallman2001, kallman2004} to derive the physical parameters for the \xray\ outflowing gas (primarily its location and the mass outflow rate), which are clearly affected by large uncertainties given the limited significance of the absorption line. 
The relatively large restframe EW of the iron absorption feature in the \xmm\ best-fitting model ($\sim210$~eV assuming $z=1.59$, see Table~\ref{xray_results}), combined with the curve of growth for highly ionized iron transitions (see Fig.~1 of \citealt{tombesi2011}), suggests the need for high turbulent velocities. For this reason, we chose, among the {\sc xstar} grids, the one with a turbulent velocity of 5000~km~s$^{-1}$. 
This value is not at odds with the line width ($\sigma$), whose 90\% upper limit is $\sim$~150~eV (equivalent to $\sim$6000~km~s$^{-1}$) in \xmm\ spectra. 
The free parameters of the {\sc xstar} model are the column density ($N_{\rm H}$) of the ionized gas, the ionization parameter (defined as $\xi=L_{\rm ion}/n\ r^{2}$, where $L_{\rm ion}$ is the ionizing radiation in the 13.6~eV--13.6~keV energy range and $nr$ the measured column density $N_{\rm H}$), and the observed redshift of the absorber $z_{\rm o}$, which is related to the intrinsic absorber redshift (i.e., in the source restframe) $z_{\rm a}$ as $(1+z_{o})=(1+z_{a})\ (1+z_{c})$, where $z_{\rm c}$=1.59. 
The velocity can then be determined using the relativistic Doppler formula 
$1+z_{a}=((1-\beta)/(1+\beta))^{0.5}$, where $\beta=v/c$ is required here to be positive for an outflow (see \citealt{tombesi2011} for further details). 

Using {\sc xstar}, we were able to obtain a relatively good fit for our data ($\chi^{2}$/d.o.f.=273.5/267), and the derived spectral parameters related to the powerlaws, cold absorption, and iron emission line are broadly consistent, within their uncertainties, with those presented in $\S$\ref{xmmcdfs}. 
The resulting column density of the ionized gas and its ionization parameter are  \Nh=$(2.23^{+3.09}_{-1.60})\times10^{23}$~cm$^{-2}$ and  $\log(\xi)=2.94^{+1.47}_{-0.38}$~erg\ cm\ s$^{-1}$, respectively. These values are in good agreement with those found in AGN with \xray\ outflows (e.g., \citealt{tombesi2010, gofford2013, chartas2014}). 
From the redshift of the absorber, $z_{0}=1.24^{+0.15}_{-0.04}$, we are able to estimate $v_{\rm out}=0.14^{+0.02}_{-0.06}$c, i.e., the absorption line is either associated with the FeXXV He$\alpha$ transition at 6.70~keV or the FeXXVI Ly$\alpha$ transition at 6.97~keV (or a combination of the two). 
Given the errors associated with the velocity of the outflow and the other spectral parameters, it is difficult to clearly determine what the exact ionization state of the line is. 
We note, however, that the 6.7~keV solution generally allows for higher EW values (i.e., more consistent with ours) at a given $\log(\xi)$ (see Fig.~2 of \citealt{tombesi2011}). 

In the following calculations, we adopt the ``prescriptions" of \cite{tombesi2012a} and  \cite{crenshaw_kraemer2012}; see also \cite{krongold2007} and \cite{pounds_reeves2009} for additional discussion of outflows and related geometry. 

The minimum distance of the outflowing gas can be estimated from the radius at which the observed velocity corresponds to the escape velocity, i.e., 
$r_{\rm min}=2GM_{\rm BH}/v^{2}_{out}$, where v$_{out}$ is the measured outflow velocity\footnote{We assume that the gas is ejected at the observed velocity, without accounting for possible additional acceleration of the gas in the flow.}. To this aim, we estimate the mass of the black hole directly from the SED fitting ($\S$\ref{sed}), where the stellar mass ($M_{\star}\approx4.9\times10^{11}$~\msun, assuming a \citealt{salpeter1959}  initial mass function) has been converted into a black hole mass adopting the relation reported in Eq.~8 of \cite{sani2011}. The rms in this relation implies an error on the derived mass of a factor of $\sim2.4$.  
From the black hole mass of $\sim5.0\times10^{8}$~\msun\,, we derive $r_{\rm min}\sim7.6\times10^{15}$~cm, which is equivalent to $\sim$100 gravitational radii 
($r_{\rm g}=GM_{\rm BH}/c^{2}$). Considering the uncertainties on the black hole mass, the range for $r_{\rm min}$ is $\approx(2.6-26)\times10^{15}$~cm. 

Using the definition of the ionization parameter, we can also estimate the maximum distance of the gas from the central source as $r_{\rm max}<L_{\rm ion}/\xi\ N_{H}$ of $\sim2\times10^{4}$ gravitational radii.\footnote{The implicit assumption in the $r_{\rm max}$ estimate is that the ionizing source is seen as pointlike by the absorber.}. At this point, we can provide a crude estimate of the mass outflow rate using the formula reported in Sect.~3.1 of \cite{crenshaw_kraemer2012}:
\begin{equation}
\dot{M}_{out}=4 \pi m_{\rm p} \mu N_{\rm H} v_{\rm out} r C_{\rm g}, 
\end{equation}
where $m_{\rm p}$ is the mass of the proton, $\mu$ the mean atomic mass per proton (=1.4 for solar abundances), $N_{\rm H}$ the column density of the ionized gas, $v_{\rm out}$ is 
assumed to correspond to the line of sight outflow velocity, $r$ is the absorber's radial location, and $C_{\rm g}$ the global covering factor. Using a sample of local Seyfert galaxies,  \cite{tombesi2010} estimate the covering factor statistically to be $\sim$0.5 (see also \citealt{gofford2013} and \citealt{crenshaw_kraemer2012}). 
In the biconical approximation of the wind, this corresponds to $C_{\rm g}=\Omega/4\pi=0.5$, where $\Omega$ is the solid angle subtended, on average, by the wind with respect to the \xray\ source. 
In the following, we assume $r$ as the minimum distance of the outflowing gas 
($r_{\rm min}$), so the derived parameters are to be considered as lower limits. 
We note, however, that our object is at least an order of magnitude more luminous than the  samples of local Seyfert galaxies from which $C_{\rm g}$ was estimated statistically. Although we do not know this value exactly at high redshift and high luminosities, we can assume that it should be of this order, since it is most likely associated with a wide-angle accretion disk wind. 

We estimate a mass outflow rate of $\dot{M}_{out}\approx1.7$~\msun~yr$^{-1}$. 
The large uncertainties in the quantities used to estimate $\dot{M}_{out}$ imply a range for the mass outflow rate of $\sim(0.3-6.8)$~\msun~yr$^{-1}$ (1$\sigma$ significance level). 
The uncertainty in the outflow rate is already quite large, and thus all quantities reported below  will have a similarly large uncertainty. 
 
As a consistency check, we computed the source accretion rate. Using the bolometric luminosity of PID352 ($\sim10^{46}$~\lum) derived from the SED fitting and ascribed to accretion processes (i.e., the mid-IR emission as due to thermally reprocessed AGN emission; see \citealt{vignali2011} for details) and assuming an energy conversion factor $\eta=0.1$, we obtain $\dot{M}_{acc}\approx1.7$~\msun~yr$^{-1}$, i.e., of the same order of magnitude as the  mass outflow rate. 

At this stage, one might consider whether the outflow mechanical energy is sufficient to influence the stellar processes in the host galaxy, following the reasoning of \cite{king2010a} and \cite{pounds2014}. Assuming an outflow velocity of 0.14c, we obtain a mechanical energy rate $\dot{E}_{\it mech}=(1/2)\ \dot{M}_{out}\ v_{out}^{2}\approx9.5\times10^{44}$~\lum\  (uncertain by at least one order of magnitude), which is only 1.5\% of the Eddington luminosity of PID352 ($\sim6.5\times10^{46}$~\lum).
Using a M$_{\rm bulge}\approx500\times\ M_{\rm BH}$ value and a dispersion velocity $\sigma\approx280$~\kms\ (as expected from the M--$\sigma$ relation 
given the mass of PID352 black hole; \citealt{gultekin2009}), the binding energy of the bulge gas, E$_{\rm bind}\sim\ M_{\rm bulge} \sigma^{2}$, is $\sim3.8\times10^{59}$~erg. In a Salpeter time, the mechanical energy injected by the wind into the surrounding galaxy medium is $\sim1.3\times10^{60}\eta_{\rm mech}$~erg, where $\eta_{\rm mech}$ is the unknown fraction of the wind energy actually transferred to the bulge gas. 
Although, at face value, the energy injected by the outflow could be even greater than the binding energy of the bulge, we note that maintaining such a wind on bulge scale for a Salpeter time appears far from being an easy process. 
We also note that the efficiency of the outflow in PID352, $(1/2)\ \dot{M}_{out}\ v_{out}^{2}/L_{bol}\sim0.09$, is consistent with the values required for efficient AGN feedback driven by winds (e.g., \citealt{dimatteo2005, hopkins_elvis2010}). As such, the wind in PID352 is likely to have a significant effect on the host galaxy. The relevance of the current result relies on the high redshift of PID352, close to the observed peak of the black hole accretion rate density (e.g., \citealt{delvecchio2014}). 
Compared to other high-redshift ($z=1.5-3.9$) quasars referred to in the literature, the efficiency of the outflow in PID352 is not so extreme (efficiencies of $\sim$ 0.1, 0.3, and 1.0 were derived for the lensed quasars HS~0810$+$2554, PG~1115$+$080 and APM~08279$+$5255, respectively; \citealt{chartas2007, chartas2014}), but our estimate is based on lower-quality \xray\ data and suffers from large uncertainties. We note that high values for this efficiency may favor magnetic driving as the acceleration mechanism for the wind (e.g., \citealt{blandford_payne1982}).
We also note that the derived wind momentum, $\dot{P}_{out}=\dot{M}_{out}\ v_{out}$, is consistent with $L_{bol}/c$.

From recent literature (e.g., \citealt{tombesi2010, gofford2013}), the median velocity derived from the highly ionized iron transitions observed in local AGN is $\sim$0.1c; a similar velocity has also been observed in highly ionized oxygen transitions in Ark~564 \citep{gupta2013}. This average value is close to the velocity derived for the outflow observed in PID352. Compared with other AGN showing UFOs, the peculiarity of PID352 is that it represents one of the few cases where an \xray\ outflowing wind has been clearly detected at $z>1$ (see also \citealt{chartas2002, chartas2007, chartas2009, chartas2014, lanzuisi2012}). Furthermore, to our knowledge, PID352 is not a lensed quasar, while most of the $z>1$ outflow detections are related to lensed systems. 
Compared to the lensed high-redshift quasars with outflows HS~0810$+$2554 (z=1.51; \citealt{chartas2014}) and H1413$+$117 (z=2.56; \citealt{chartas2007}), the bolometric luminosity of PID352 is a factor of $\sim$3 and $\sim(5-10)$ higher.\footnote{We 
used a bolometric correction of 25 to convert the 2--10~keV luminosity into a bolometric luminosity, and magnification factors of 94 and (20--40) to have intrinsic values for HS~0810$+$2554 and H1413$+$117, respectively.} 
The relatively good constraints for the outflow parameters of PID352 are mostly due to the extensive observational \xray\ campaign carried out in the CDF-S over more than a decade (in this regard, see also the claimed but never confirmed outflows reported by \citealt{wang2005, zheng_wang2008}) and by the large collecting area of \xmm. 

As a final remark, we briefly discuss the presence of the wind in PID352 and its possible link to radio jets, since the source is classified as an FRII galaxy (see $\S$\ref{radio}). As discussed by \cite{tombesi2014}, where a minimum wind covering fraction of $C\approx(0.3-0.7)$ was derived, UFOs in radio-loud AGN are likely to be covering a significant portion of the sky as seen by the central source. Since derived statistically for a sample of 26 radio-loud AGN (using also the information from the radio jet inclination angle to visualize the geometry of the system, assuming that the jet is perpendicular to the accretion disk), this result implies that the absorbing material is not highly collimated as expected in the case of jets; as a result, an accretion-disk wind is favored. Furthermore, in the specific case of another FRII radio galaxy, 3C~111, the UFO mass outflow rate ($\sim0.1-1$~\msun~yr$^{-1}$, i.e., similar to that of PID352) is much greater than the outflow funnelled into the jet ($\dot{M}_{out,j}=0.0005-0.005$~\msun~yr$^{-1}$; \citealt{tombesi2012b}), implying that the UFO is much more massive than the jet, although its kinetic power can be one order of magnitude lower. Placed in a broader context, these results are suggestive of considerable feedback onto the host galaxy by winds and jets 
(see also the discussion in \citealt{miller2009} about the co-existence of outflows and jets in radio-loud quasars and their connection with \xray\ binaries).

\section{Summary}
\label{summary}
In this work we have investigated the properties of PID352, a luminous (\lx$\approx4\times10^{44}$~erg/s), obscured (\Nh$\approx2\times10^{23}$~cm$^{-2}$) quasar at $z_{X}\approx1.6$ detected in the CDF-S with both \xmm\ and \chandra\ and associated with the radio-undetected core of an FRII galaxy at 20cm, which is optically classified as an extremely red object. 
\xray\ spectral data show the presence of an iron-line complex consisting of an emission  plus an absorption line. While the former feature (used, in conjunction with the iron absorption edge, to estimate the source redshift) is probably associated with neutral iron K$\alpha$ emission, the latter, according to a fit with {\sc xstar}, is ascribed either to the FeXXV or to the FeXXVI transition (or a combination of the two) in an outflowing gas, implying $v_{\rm out}=0.14^{+0.02}_{-0.06}$c. 
The outflow velocity is within the observed range for local Seyfert galaxies (e.g., \citealt{tombesi2010, gofford2013}). 
Monte Carlo simulations indicate that the probability of detecting an absorption line when it does not exist and providing an improvement to the spectral fitting larger or comparable to the one we obtain is $\sim$1\%. 
A basic and rather uncertain calculation places the minimum distance of the outflowing gas at $\sim$~100~gravitational radii from the black hole, i.e., on scales typical of the accretion disk. In the panorama of AGN with high-velocity outflowing gas, the peculiarity of PID352 consists in being one of the few sources with a detected UFO at $z>1$ without being a lensed quasar. 

For PID352 a mass outflow rate of $\sim$2~\msun~yr$^{-1}$ (range 0.3--6.8~\msun~yr$^{-1}$) is derived, which is consistent with the source accretion rate. 
We have also estimated that the mechanical power injected by the wind onto the host galaxy may be very high and potentially able to disrupt the bulge, provided that a large amount of the wind energy is efficiently transferred to the bulge gas on long ($\gtrsim$ Salpeter) timescales, which looks like an extreme scenario.  In any case, the wind is likely to have a significant impact on the host galaxy. 

An interesting step forward in the study of PID352 can consist of detecting large-scale outflows at longer wavelengths. This would allow us to investigate the physical link between the small-scale high-velocity outflow probed by X-rays and the outer-scale outflows \citep{tombesi2015}. In particular, given the source magnitude in the near-IR (AB magnitude $\sim$21.8 and 21.3 in the {\it J} and {\it H} bands, respectively), a moderate-length ($\sim$1~hr) observation at ESO with VLT/XShooter (allowing for sensitive, medium-resolution spectroscopy up to the {\it K}  band) will be sufficient to detect the main restframe optical emission lines (H$\beta$, \oiii,\ and H$\alpha$) at a signal-to-noise ratio of $\sim$10 and derive a more accurate measure of the redshift; possibly, this observation will also infer the presence of kpc-scale outflow in the ionized gas  component. 
A more efficient but more time-consuming way to detect the outflow on kpc scales would be to use VLT/SINFONI in the {\it J} and {\it H} bands. 
Moving to sub-mm wavelengths, \alma\ observations will allow us to extend these studies to the molecular component and probe the link between UFOs and large-scale outflows (e.g., \citealt{tombesi2015, feruglio2015}).

\begin{acknowledgements}
The authors thank the referee for detailed and thoughtful comments and suggestions. 
Financial contribution from ``PRIN--INAF 2011" and ``PRIN--INAF 2012" is acknowledged. 
KI acknowledges support by the Spanish MINECO under grant AYA2013-47447-C3-2-P 
and MDM-2014-0369 of ICCUB (Unidad de Excelencia ``Mar\'\i{}a de Maeztu"). 
GL and MB acknowledge support from the FP7 Career Integration Grant ``eEASy"' (``SMBH evolution through cosmic time: from current surveys to eROSITA-Euclid AGN Synergies", CIG 321913).
FJC acknowledges Financial support from the Spanish Ministerio de Econom\'\i{}a y Competitividad under project AYA2012-31447. 
WNB thanks support from Chandra X-ray Center grant G04-15130A and NASA ADP grant NNX10AC99G. 
FEB acknowledges support from CONICYT-Chile (Basal-CATA PFB-06/2007, FONDECYT  1141218, ``EMBIGGEN" Anillo ACT1101) and the Ministry of Economy, Development, and  Tourism's Millennium Science Initiative through grant IC120009, awarded to The Millennium Institute of Astrophysics, MAS.
CV thanks P. Ciliegi, D. Dallacasa and E. Middelberg for information on radio data, and M. Cappi, G. Chartas, A. Feltre, P. Grandi, E. Lusso, G. Ponti, L. Pozzetti, E. Rovilos, P. Severgnini, F. Vito, and G. Zamorani for useful discussions. 
\end{acknowledgements}

%-------------------------------------------------------------------

\end{document}